\documentclass[12pt]{iopart}
\usepackage{graphicx}
\usepackage{caption}
\begin{document}
\title{Partonic effects on anisotropic flows at RHIC}

\author{Lie-Wen Chen
\footnote[1]{On leave from Department of Physics, Shanghai Jiao Tong 
University, Shanghai 200030, China}\footnote[2]{lwchen@comp.tamu.edu}
and Che Ming Ko}

\address{Cyclotron Institute and Physics Department, Texas A\&M
University, College Station, Texas 77843-3366}

\begin{abstract}
We report recent results from a multiphase transport (AMPT) model on
the azimuthal anisotropies of particle momentum distributions in heavy
ion collisions at the Relativistic Heavy Ion Collider. These include
higher-order anisotropic flows and their scaling, the rapidity
dependence of anisotropic flows, and the elliptic flow of charm quarks.
\end{abstract}

%\submitted

% \maketitle

\section{Introduction}

Anisotropic flows in heavy ion collisions at the Relativistic Heavy
Ion Collider (\textrm{RHIC}) are sensitive to the properties of
produced matter. This sensitivity
not only exists in the larger elliptic flow 
\cite{tean,kolb,Zhang99,moln1,Lin:2001zk,gyulv2,Voloshin03} but also in the
smaller higher-order anisotropic flows \cite{Kolb03,chen04,Kolb04}.
Experimentally, scaling relations among hadron anisotropic flows
have been observed \cite{STAR03}, and according to the quark
coalescence model they are related to similar scaling relations among
parton anisotropic flows \cite{Kolb04}. Also, anisotropic flows 
measured at finite pseudorapidities are seen to depend strongly on the
rapidity \cite{STAR03,phobos02,manly03,tonjes04,oldenburg04,back04}, and
this has so far not been reproduced by theoretical models \cite%
{hirano01,heinz04,csanad04,chen04v13}.  For charmed mesons, the observed
transverse momentum spectra of electrons from their decays are
found to be consistent with both limiting scenarios of perturbative
QCD spectra without final-state interactions and complete
thermalization including transverse expansion \cite{Batsouli03}. The charm
elliptic flows given by the quark coalescence model show, however, marked
differences between these two scenarios \cite{Greco04}. These interesting
phenomena have recently been studied in a multiphase transport (AMPT)
model that includes both initial partonic and final hadronic
interactions \cite{Zhang:2000bd,Lin:2001cx,LinHBT02,ko,zhang,pal}.
In this talk, we briefly review the AMPT model and discuss the results
obtained from this model.

\section{The AMPT model}

The \textrm{AMPT} model is a hybrid model that uses minijet partons
from hard processes and strings from soft processes in the Heavy Ion
Jet Interaction Generator (\textrm{HIJING}) model \cite{Wang:1991ht}
as the initial conditions for modeling heavy ion collisions at
ultra-relativistic energies. Time evolution of resulting minijet
partons is then described by Zhang's parton cascade (\textrm{ZPC})
\cite{Zhang:1997ej} model. At present, this model includes only
parton-parton elastic scatterings with an in-medium cross section
derived from the lowest-order Born diagram with an effective gluon
screening mass taken as a parameter for fixing the magnitude and angular
distribution of parton scattering cross section. After
minijet partons stop interacting, they are combined with their parent 
strings, as in the \textrm{HIJING} model with jet quenching, to
fragment into hadrons using the Lund string fragmentation model as
implemented in the \textrm{PYTHIA} program \cite{Sjostrand:1994yb}. The
final-state hadronic scatterings are then modeled by a relativistic
transport (\textrm{ART}) model \cite{Li:1995pr}.

Since the initial energy density in Au + Au collisions at
\textrm{RHIC} is much larger than the critical energy density at which
the hadronic matter to quark-gluon plasma transition would occur
\cite{zhang,Kharzeev:2001ph}, the \textrm{AMPT} model has been
extended by converting the initial excited strings into partons 
\cite{Lin:2001zk}. In this string melting scenario, hadrons, that would have
been produced from string fragmentation, are converted instead to valence
quarks and/or antiquarks. Interactions among these partons are again
described by the\textrm{\ ZPC} parton cascade model. Since there are no
inelastic scatterings, only quarks and antiquarks from the melted strings
are present in the partonic matter. The transition from the partonic
matter to the hadronic matter is achieved using a simple coalescence
model, which combines two nearest quark and antiquark into mesons and
three nearest quarks or antiquarks into baryons or anti-baryons that
are close to the invariant mass of these partons. The present
coalescence model is thus somewhat different from the ones recently
used extensively \cite{greco,hwa,fries,molnar03} for studying hadron
production at intermediate transverse momenta. 

\section{Anisotropic flows at midrapidity}

\begin{figure}[htb]
\begin{minipage}{20pc}
\begin{center}
\includegraphics[scale=0.9]{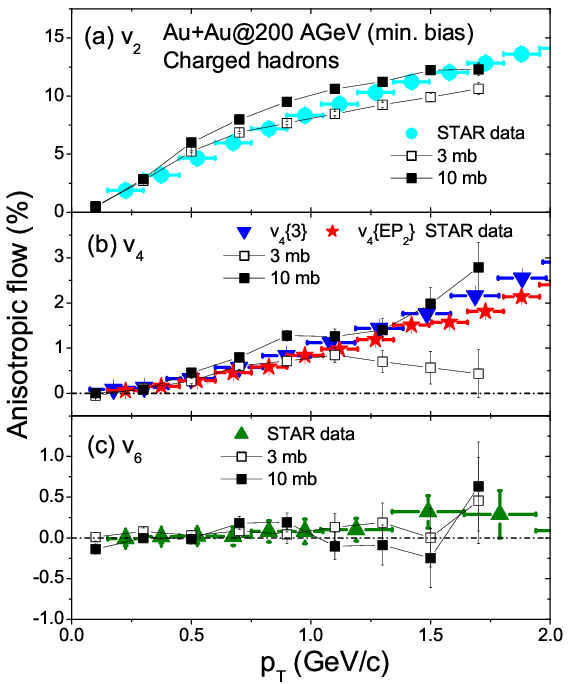} 
\caption{{\protect\small Anisotropic flows }$v_{2}$%
{\protect\small \ (a), }$v_{4}${\protect\small \ (b), and }$v_{6}$%
{\protect\small \ (c) of charged hadrons in} $%
\left\vert \protect\eta \right\vert <1.2${\protect\small \ from minimum bias
Au + Au collisions at }$\protect\sqrt{s}=200${\protect\small \ \textrm{AGeV}
as functions of }$p_{T}$ {\protect\small \ for parton
cross sections of } $3${\protect\small \ (open
squares) and } $10${\protect\small \ (solid squares) \textrm{mb}. 
Data are from the STAR Collaboration \protect\cite{STAR03}.}
\label{v246ptchg}}
\end{center}
\end{minipage}\hspace{0pc}%
\begin{minipage}{19pc}
\begin{center}
\includegraphics[scale=0.8]{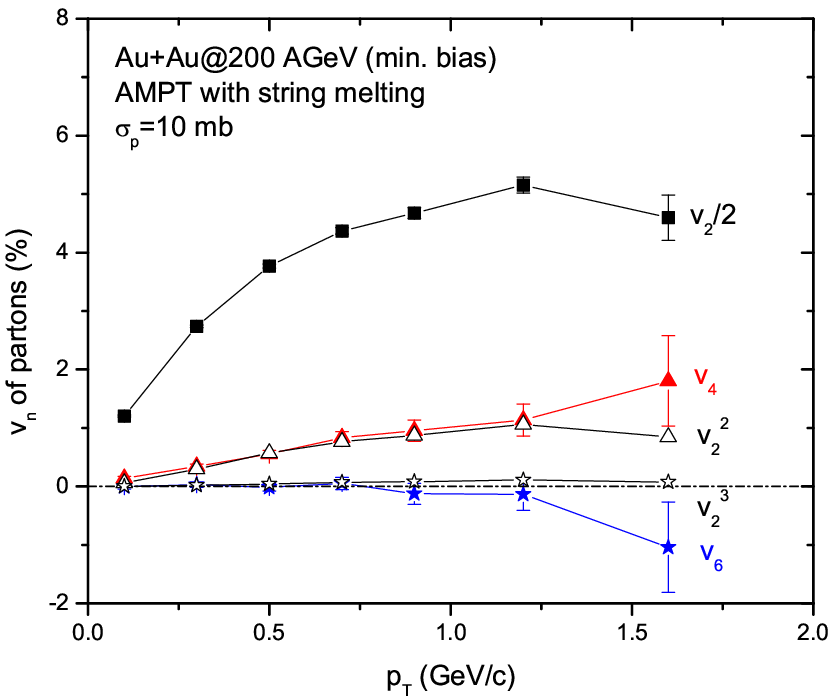} 
\caption{{\protect\small \ Transverse
momentum dependence of midrapidity parton anisotropic flows }$v_{2}$%
{\protect\small , }$v_{4}${\protect\small \ and }$v_{6}${\protect\small \
from minimum bias events for Au + Au at }$\protect\sqrt{s}=200$%
{\protect\small \ \textrm{AGeV} with parton scattering cross section }$10$%
{\protect\small \ \textrm{mb}. Also plotted are }$v_{2}^{2}${\protect\small %
\ (open triangles) and }$v_{2}^{3}${\protect\small \ (open stars).}
\label{v246ptparton}}
\end{center}
\end{minipage} 
\end{figure}

Using the \textrm{AMPT} model with string melting, we have studied
anisotropic flows in the pseudorapidity range 
$\left\vert \eta \right\vert <1.2$ in minimum bias Au + Au collisions
at $\sqrt{s}=200$ \textrm{AGeV}. In Fig. \ref{v246ptchg}, we show
the final anisotropic flows $v_{n}$ of charged hadrons, defined by the
average $<\cos(n\phi)>$ of the azimuthal distributions of their
transverse momenta, as functions of transverse momentum $p_{T}$ for
parton scattering cross sections $\sigma _{p}=3$ and $10 $ \textrm{mb},
together with recent experimental data from the \textrm{STAR}
collaboration \cite{STAR03}. It is seen that the parton scattering
cross section $\sigma _{p}=3$ \textrm{mb} underestimates the data at
higher $p_{T}$ ($>1$ \textrm{GeV/}$c$) while $\sigma _{p}=10$
\textrm{mb} seems to give a better fit to the data. The values of hadronic $%
v_{6}$ are in agreement with the data within error bars, although they are
essentially zero. The $v_{4}$ of charged hadrons exhibits a stronger
sensitivity to the parton cross section than their $v_{2}$, especially at
higher $p_{T}$, and is thus a more sensitive probe to the initial
partonic dynamics in relativistic heavy ion collisions.

The experimental data shown in Fig. \ref{v246ptchg} on higher-order
anisotropic flows show the scaling relation
$v_{n}(p_{T})\sim v_{2}^{n/2}(p_{T})$ \cite{STAR03}. In the naive
quark coalescence model \cite{molnar03} that only allows quarks with
equal momentum to form a hadron, meson higher-order anisotropic flows
$v_{4,M}$ and $v_{6,M}$ are related to parton anisotropic flows $v_{n,q}$
by \cite{kolb,chen04,Kolb04}
\begin{equation}
\hspace{-1.3cm}
\frac{v_{4,M}(p_{T})}{v_{2,M}^{2}(p_{T})}\approx \frac{1}{4}+\frac{1}{2}%
\frac{v_{4,q}(p_{T}/2)}{v_{2,q}^{2}(p_{T}/2)},~~~\allowbreak \frac{%
v_{6,M}(p_{T})}{v_{2,M}^{3}(p_{T})}\approx \allowbreak \frac{1}{4}\left(
\frac{v_{4,q}(p_{T}/2)}{v_{2,q}^{2}(p_{T}/2)}+\frac{v_{6,q}(p_{T}/2)}{%
v_{2,q}^{3}(p_{T}/2)}\right).  \label{v6Mscal}
\end{equation}%
The meson higher-order anisotropic flows thus satisfy scaling
relations if such relations exist among quark higher-order
anisotropic flows.

In Fig. \ref{v246ptparton}, we show the $p_{T}$ dependence of the
anisotropic flows $v_{2}$, $v_{4}$ and $v_{6}$ of midrapidity partons
obtained from the \textrm{AMPT} model with string melting and a parton
scattering cross section of $10$ \textrm{mb} for above
reaction. Also shown in Fig. \ref{v246ptparton} are $v_{2}^{2}$ (open
triangles) and $v_{2}^{3}$ (open stars). Comparing them to $v_4$ and
$v_6$, respectively, indeed show that the parton anisotropic flows
satisfy the scaling relation $v_{n,q}(p_{T})\sim v_{2,q}^{n/2}(p_{T})$.
With a parton scaling factor of $1$, the naive coalescence model would
lead to the following scaling relations for meson anisotropic flows:
\begin{equation}
\frac{v_{4,M}(p_{T})}{v_{2,M}^{2}(p_{T})}\approx \frac{3}{4},~~~\frac{%
v_{6,M}(p_{T})}{v_{2,M}^{3}(p_{T})}\approx \allowbreak \frac{1}{2}.
\label{v6MscalS}
\end{equation}
The resulting hadron scaling factors of $3/4$ and $1/2$ are, however,
smaller than those ($\sim 1.2$) extracted from measured anisotropic
flows of charged hadrons. Since the naive quark coalescence model does
not allow hadron formation from quarks with different momenta as in
more realistic quark coalescence models \cite{greco,hwa,fries}, it is
not expected to give a quantitative description of the experimental
observation. Such effects are, nevertheless, included in the
\textrm{AMPT} model, which have been shown in Fig. \ref{v246ptchg} to
reproduce measured hadron anisotropic flows.

\section{Pseudorapidity dependence of anisotropic flows}

\begin{figure}[htb]
\begin{minipage}{19pc}
\begin{center}
\includegraphics[scale=0.8]{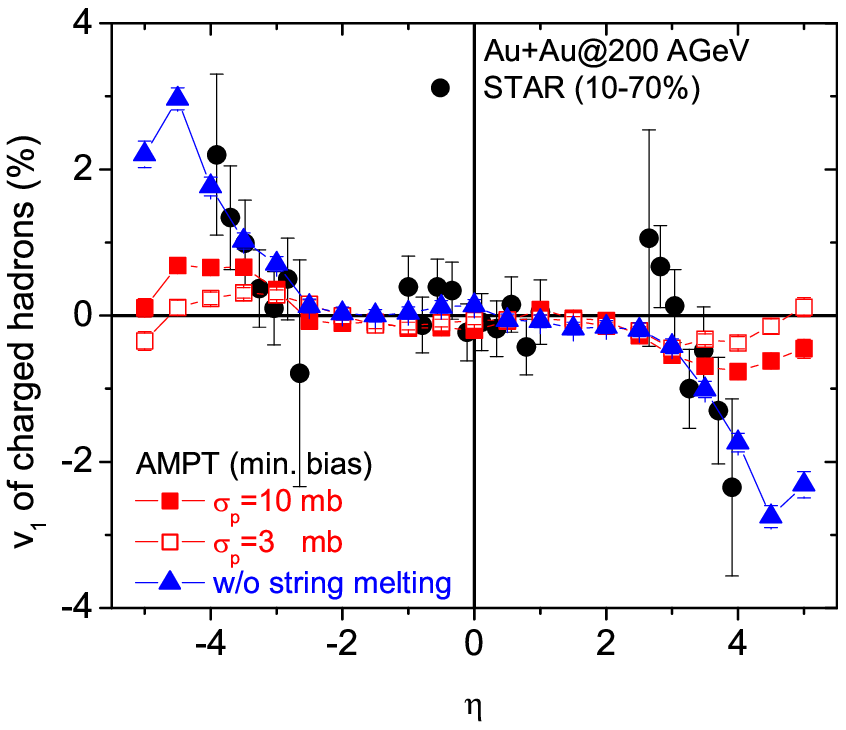}
\caption{{\protect\small Pseudorapidity dependence of }$v_{1}$%
{\protect\small \ from minimum bias events of Au + Au collisions at }$%
\protect\sqrt{s}=200${\protect\small \ AGeV in the string melting scenario
with parton scattering cross sections }$\protect\sigma _{p}=3$%
{\protect\small \ (open squares) and }$10${\protect\small \ (solid squares)
\ mb as well as the scenario without string melting (solid triangles). Data
are from the STAR collaboration \protect\cite{STAR03}.}\label{v1yh}}
\end{center}
\end{minipage}\hspace{0pc}%
\begin{minipage}{20pc}
\begin{center}
\includegraphics[scale=0.8]{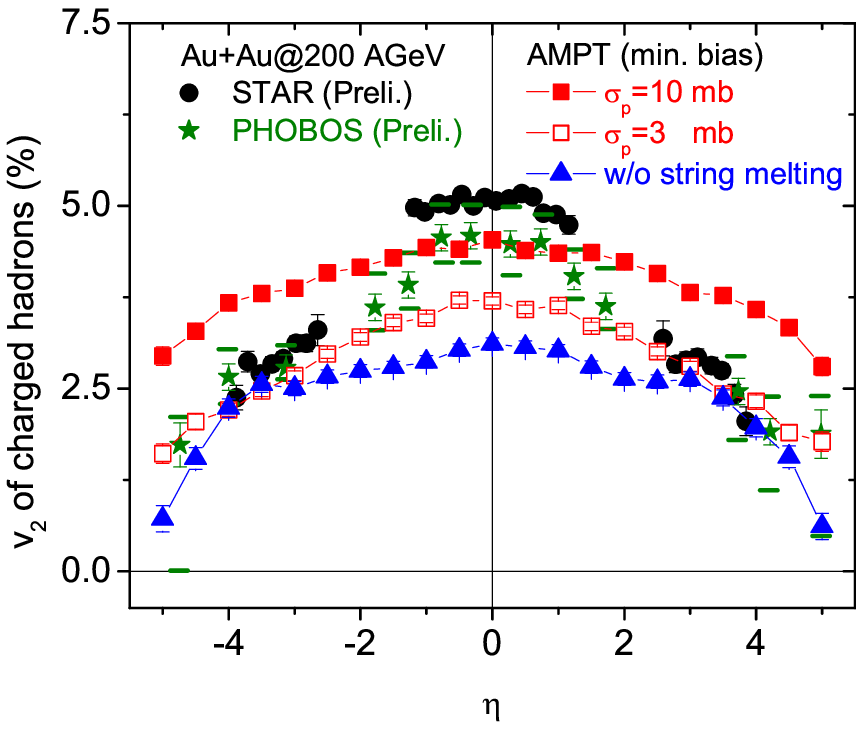}
\caption{{\protect\small Pseudorapidity dependence of }$v_{2}$%
{\protect\small \ from minimum bias
Au + Au collisions at }$\protect\sqrt{s}=200${\protect\small \ AGeV in the
string melting scenario with parton cross sections of}
$3${\protect\small \ (open squares) and }$10${\protect\small \
(solid squares) \ mb and the scenario without string melting (solid
triangles). Data are from the PHOBOS (stars)\protect\cite{manly03} and STAR
collaborations (solid circles) \protect\cite{oldenburg04}.}\label{v2yh}}
\end{center}
\end{minipage} 
\end{figure}

Results from the AMPT model on the pseudorapidity dependence of
$v_{1}$ for charged hadrons from minimum bias events of Au + Au
collisions at $\sqrt{s}=200$ \textrm{AGeV} are shown in 
Fig. \ref{v1yh} for the scenarios of string melting with parton
scattering cross sections $\sigma _{p}=3$ (open squares) and $10$ \textrm{mb}
(solid squares) as well as for the scenario without string melting
(solid triangles). Also included in Fig. \ref{v1yh} are recent data
from the \textrm{STAR} collaboration (solid circles)
\cite{STAR03}. Both scenarios can reproduce approximately the data
around the mid-pseudorapidity region, i.e., $v_{1}$ is flat
(essentially zero) around mid-$\eta $. For $v_{1}$ at large 
$\left\vert \eta \right\vert $, the string melting scenario with both
parton scattering cross sections $\sigma _{p}=3$ \textrm{mb} and $10$ 
\textrm{mb}\ underestimates significantly the data. On the other hand, the
scenario without string melting seems to give a good description of $v_{1}$
at large $\left\vert \eta \right\vert $.

The predicted pseudorapidity dependence of charged hadron $v_{2}$ from the
same reaction is shown in Fig. \ref{v2yh}, together with preliminary data
from the \textrm{PHOBOS} collaboration (solid stars) \cite{manly03} and the 
\textrm{STAR} collaboration (solid circles) \cite{oldenburg04}. The
string melting scenario with $\sigma _{p}=$ $10$ \textrm{mb} (solid
squared) is seen to describe very well the data for $v_{2}$ around
mid-$\eta $ ($\left\vert \eta \right\vert \leq 1.5$) but
overestimates the data at large pseudorapidity.  The overestimation of
may stem from the use of constant parton scattering cross section in 
the AMPT model. Since the properties of partonic matter at different
rapidities may not be the same in heavy ion collisions at \textrm{RHIC},
different parton cross sections may have to be used. Comparison between
theoretical results and the experimental data for elliptic flow indicates
that a larger $\sigma _{p}=$ $10$ \textrm{mb} is needed at midrapidity but a
smaller $\sigma _{p}=$ $3$ \textrm{mb} (open squares) gives a better
description at large pseudorapidity. Also shown in Fig. \ref{v2yh} are
results obtained from the scenario without string melting (solid triangles),
and they are seen to also describe the data at large pseudorapidity ($%
\left\vert \eta \right\vert \geq 3$). Therefore, the scenario without
string melting can describe simultaneously the data on $v_{1}$ and $v_{2}$
at large pseudorapidity ($\left\vert \eta \right\vert \geq 3$). These
interesting features imply that initially the matter produced at large
pseudorapidity ($\left\vert \eta \right\vert \geq 3$) is dominated by
strings while that produced around mid-rapidity ($\left\vert \eta
\right\vert \leq 3$) mainly consists of partons. This is a reasonable
picture as particles at large rapidity are produced later in time when
the volume of the system is large and the energy density is small.

\section{Charm flows at RHIC}

\begin{figure}[th]
\begin{center}
\includegraphics[scale=0.9]{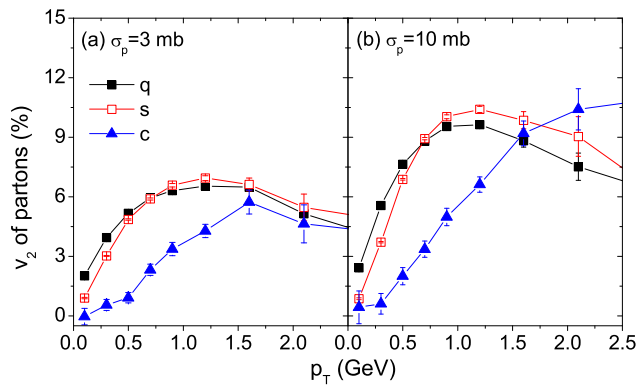}
\vspace{-0.5cm}
\caption{{\protect\small Transverse momentum dependence of $v_{2}$ for
different parton flavors in midrapidity from minimum bias Au + Au
collisions at $\sqrt{s}=200$ \textrm{AGeV}. Results are for parton
cross sections $\sigma _{p}=3$ (a) and $10$ \textrm{mb} (b). }}
\label{v2dsc}
\end{center}
\end{figure}

Using the \textrm{AMPT} model in the string melting scenario with same
parton scattering cross section of $\sigma _{p}=3$ or $10$
\textrm{mb} for all quarks, we have also studied the
$p_{T}$-dependence of $v_{2}$ for partons of different flavors in
the midrapidity from minimum bias Au + Au collisions at
$\sqrt{s}=200$ \textrm{AGeV}. Using the current quark mass of $10$ 
\textrm{MeV} for \textsl{d} quark, $6$ \textrm{MeV} for
\textsl{u} quark, $200$ \textrm{MeV} for \textsl{s} quark, and
$1.35$ \textrm{GeV} for \textsl{c} quark, the results are shown in
Fig. \ref{v2dsc}.  It is seen that the quark $v_2$ increases with
increasing $p_T$, and at $p_T<1.5$ GeV/c it is smaller for the heavier 
$c$ quark than for the lighter $d$ and $s$ quarks. This mass
dependence of $v_{2}$ at low $p_T$ is similar to the mass ordering of
hadron elliptic flows in the hydrodynamic model which assumes
that the matter is in local thermal equilibrium and thus develops a large
collective radial flow. However, instead of continuing increase of
$v_2$ with respect to $p_T$ as in the hydrodynamic model, 
the parton $v_2$ in the transport model reaches a maximum value
at certain large $p_T$, indicating that high momentum particles do not 
achieve thermal equilibrium with the bulk matter. These features 
are seen for both values of parton scattering cross sections, except that
the value of $v_2$ is larger for a larger parton scattering cross
section. As in the case of light hadrons, study of charmed hadron
flows offers the possibility to understand the charm quark
interactions in the quark-gluon plasma. 

\section{Summary and discussions}

In summary, using the \textrm{AMPT} model, we have studied the partonic
effects on anisotropic flows in heavy ion collisions at \textrm{RHIC}. We
find that measured $v_{2}$, $v_{4}$ and $v_{6}$ of charged hadrons at
midrapidity in Au + Au collisions at $\sqrt{s}=200$ AGeV can be described
by a parton scattering cross section of about $10$ mb and that $v_{4}$
is a more sensitive probe to the initial partonic dynamics in these
collisions than $v_{2}$. Moreover, higher-order parton anisotropic
flows are nonnegligible and satisfy the scaling relation 
$v_{n,q}(p_{T})\sim v_{2,q}^{n/2}(p_{T})$, which leads naturally to
the observed similar scaling relation among hadron anisotropic flows
if the partonic matter hadronizes via the coalescence of quarks and
antiquarks. The results on the rapidity dependence of anisotropic flows
suggest that a partonic matter is formed during early stage of relativistic
heavy ion collisions only around midrapidity and that strings remain
dominant at large rapidities. We have also studied the elliptic flow of 
charm quarks by using different scattering cross sections, and find
that it has a maximum value similar to that for light quarks and
is also sensitive to the charm quark scattering cross sections. The 
charm quark elliptic flow shows, on the other hand, a very different
dependence on the transverse momentum due to its large mass. 

In reproducing the experimental data on hadron anisotropic flows, 
a large parton cross section of about 10 mb is needed in the AMPT model. 
Comparing to the cross section given by the perturbative QCD, this value
is an order of magnitude larger, indicating that nonperterbative effects
are important in the quark-gluon plasma. Indeed, recent lattice gauge studies
have shown that hadron spectral functions including those consisting
of heavy charm quarks survive in the quark-gluon plasma 
\cite{datta,asakawa}. Including the effect of these quasi bound states
is expected to enhance the parton-parton scattering cross section. 

\section*{Acknowledgment}

We thank V. Greco, P. F. Kolb, Z. W. Lin, and B. Zhang for
collaboration on different parts of the presented work. This paper was
based on work supported by the US National Science Foundation under Grant 
No. PHY-0098805 and the Welch Foundation under Grant No. A-1358.
LWC was also supported by the National Science Foundation of China
under Grant No. 10105008.

\end{document}